%% file: main.tex
\documentclass[12pt,a4paper]{article}
\usepackage[T1]{fontenc}
\usepackage{geometry}
\usepackage[english]{babel}
\usepackage{amsmath}
\usepackage{booktabs}
\usepackage{graphicx}
\usepackage{float}
\usepackage{placeins}
\usepackage{amsmath}
\usepackage{amssymb}
\usepackage{ae,aecompl}
\usepackage[multiple]{footmisc}
\usepackage[hyperfootnotes=false]{hyperref}
\usepackage{bookmark}
\hypersetup{colorlinks=true,citecolor=blue}
\usepackage{amsthm}
\usepackage{MJOARTI}
\usepackage{setspace}
\usepackage{graphicx}
\usepackage{rotating}
\usepackage{bbold}
\usepackage{lscape}
\usepackage{subcaption}
\usepackage{caption}
\usepackage{hyperref}
\usepackage{lscape}
\usepackage{ifthen}
\usepackage{url}
\usepackage{blkarray}
\usepackage{multirow}
\usepackage{tabularx}
\usepackage{longtable}
\usepackage{enumerate}
\usepackage{supertabular}
\usepackage{fancyhdr}
\usepackage{epstopdf}
\usepackage{dsfont}
\usepackage{url}
\usepackage{multicol}
\usepackage{titlesec}
\usepackage{eurosym}
\usepackage{arydshln}
\usepackage{natbib}
\usepackage{threeparttable} 
\usepackage{comment}

\usepackage{array}
\newcolumntype{k}{>{\centering\arraybackslash}p{2cm}}
\usepackage{longtable}
\usepackage{ltcaption}
\usepackage{multicol}
\usepackage{color}
\usepackage{textcomp}
\definecolor{dark-red}{rgb}{0.6,0,0}
\definecolor{listinggray}{gray}{0.9}
\definecolor{darkgreen}{rgb}{0,0.4,0}
\definecolor{lbcolor}{rgb}{0.9,0.9,0.9}
\usepackage{adjustbox}
\usepackage{lscape}

\begin{document}

\date{\today}    

\title{
\textbf{Unveiling Plant-Product Productivity via First-Order Conditions: Robust Replication of Orr (2022)}%
  \thanks{
Authors: 
Hong: National Taiwan University. E-mail: \href{mailto:jkhong@ntu.edu.tw}{jkhong@ntu.edu.tw}. Address: {Department of Economics, National Taiwan University, No.1, Sec 4, Roosevelt Road, Taipei, 10617 Taiwan}; Luparello: The Pennsylvania State University and BAFFI. E-mail: \href{dluparello@psu.edu}{dluparello@psu.edu}. Address: {Department of Economics, 303 Kern Building, The Pennsylvania State University, University Park, PA 16802}. We thank the editor, Heather Anderson, and two anonymous referees for their insightful feedback. This paper builds upon the report we prepared for the Institute for Replication (\cite{brodeur2024mass}), titled ``Robustness Report on `Within-Firm Productivity Dispersion: Estimates and Implications,' by Scott Orr (2022).'' We express sincere gratitude to Scott Orr, whose thoughtful suggestions substantially enhance the quality of our work. We remain equally indebted to Abel Brodeur for his support throughout this project. We also acknowledge the Ministry of Statistics and Programme Implementation, Government of India, for providing the data this project employs. {Finally, we acknowledge the research support of OpenAI’s ChatGPT and Anthropic’s Claude LLMs, used within the guidelines outlined in \cite{korinek2023generative}.}}
}
\author{Joonkyo Hong \and Davide Luparello}

\maketitle
\thispagestyle{empty}

\renewcommand{\abstractname}{Abstract}
\begin{abstract}

We assess the replicability of \cite{orr2022within}'s method for estimating within-plant productivity across product lines, which combines demand estimation with cost minimization. The original study uses input price shocks in other output markets as instrumental variables, with exclusion restrictions based on downstream purchase shares. Reconstructing the original dataset of Indian machinery producers from 2000--2007, we reproduce the main productivity patterns and demonstrate their robustness to variations in the exclusion threshold. The main results remain robust in extended samples (2010--2019, 2000--2019) when calibrating demand parameters to \cite{orr2022within}'s 2000-2007 estimates, as estimation on these extended periods yields inadmissible demand systems.\\
\newline

\textsc{Keywords}: Replication, Robustness, Productivity \newline

\textsc{JEL codes}: D22, D24, L64
\end{abstract}

\clearpage

\begin{spacing}{2}
\setcounter{page}{1}
 
\section{Introduction}

Productivity variation across product lines is a critical determinant of overall efficiency \citep{eckel2010multi, bernard2010multiple, mayer2021product}, yet empirical progress has been hampered by the unavailability of producer-product-level input data. \citet{orr2022within} addresses this challenge by developing a two-stage estimation procedure: the first stage recovers product-specific input allocations by combining the estimation of product demand systems and cost minimization conditions, while the second stage estimates the product-level production function. Analyzing data on Indian machinery producers from 2000--2007, \citet{orr2022within} establishes two main findings: eliminating the lowest-TFPR product increases plant-level TFPR by 10--65 percent, and a one standard deviation decrease in product-level TFPR raises the probability of product discontinuation by 6 percentage points.

\citet{orr2022within}'s methodology confronts a well-known endogeneity concern: product prices may correlate with demand through unobservable characteristics such as quality and appeal. To address it, \citet{orr2022within} constructs instrumental variables using average input price growth facing plants in non-machinery industries, under the identifying assumption that these input price shocks remain orthogonal to machinery industry demand. This assumption fails when machinery producers constitute a substantial share of downstream demand for a given input. To preserve instrument validity, \citet{orr2022within} excludes input codes from single-product producers when machinery-related plants account for more than 30 percent of purchases of that specific input code.

In this paper, we first reproduce the original results. Since the raw administrative data were not included in the replication package, we obtained them independently from the Indian Ministry of Statistics and Programme Implementation (MOSPI). Our estimates closely match the published results, with only minor numerical discrepancies that do not affect the study's main conclusions. Next, we conduct a sensitivity analysis of the IV construction threshold (30 percent). We find that instrument strength and estimated demand parameters are sensitive to threshold values. Nonetheless, the main results remain remarkably stable. Finally, we extend the analysis to later years (2010--2019) and the full sample (2000--2019). Across all IV thresholds, the proposed instruments yield inadmissible demand systems. We therefore calibrate the demand parameters to \citet{orr2022within}'s original estimates to recover within-plant productivity measures. The main findings remain robust.

The paper proceeds as follows. Section \ref{s:idea} outlines the main methodological innovation in \citet{orr2022within}. Section \ref{s:comp} presents our computational reproduction. Section \ref{s:IV} investigates the sensitivity of demand estimates to IV construction thresholds. Section \ref{s:post} extends the analysis to later periods. Section \ref{s:conc} concludes.

\section{Central Idea: The Use of First Order Conditions}\label{s:idea}

In the literature, unit-level productivity is often recovered as production function estimation residuals. To estimate plant-product level productivity, \citet{orr2022within} employs the following production function\footnote{{Specified in equation (18) of his original paper}.}:  
\begin{equation}
    Y_{it}^{j} = \exp(\omega_{it}^{j}) (L_{it}^{j})^{\beta_{L}} (K_{it}^{j})^{\beta_{K}} (M_{it}^{j})^{\beta_{M}},
\end{equation}
where \( Y_{it}^{j} \) is the total output of product \( j \) by plant \( i \) in period \( t \); \( L_{it}^{j} \), \( K_{it}^{j} \), and \( M_{it}^{j} \) are the respective labor, capital, and material inputs used to produce product \( j \); and \( \omega_{it}^{j} \) represents total factor productivity for producing product \( j \) (TFPQ).

This production function approach requires input allocations across product lines \((L_{it}^{j}, K_{it}^{j}, M_{it}^{j})\), yet datasets typically provide only plant-level inputs \((L_{it}, K_{it}, M_{it})\). \citet{orr2022within} resolves this limitation by exploiting the first-order conditions of cost minimization to infer unobserved input shares across products. This approach yields the following fundamental relationship\footnote{Formalized in equation (7) of the original paper.}:  
\begin{equation}
   S_{it}^{j} = \frac{MC_{it}^{j} Y_{it}^{j}}{\sum_{j \in \mathbb{Y}_{it}} MC_{it}^{j} Y_{it}^{j}}, \label{eq:input_allocation} 
\end{equation}
where \(S_{it}^{j}\) denotes the input share for product \(j\), \(Y_{it}^{j}\) represents the output of product \(j\), \(MC_{it}^{j}\) captures the marginal cost of producing \(j\), and \(\mathbb{Y}_{it}\) defines the set of products produced.

With \( Y_{it}^{j} \) observable in the data, equation \eqref{eq:input_allocation} enables product-line estimation of the production function, permitting recovery of plant-product productivity conditional on knowledge of \( MC_{it}^{j} \). Calculating \( MC_{it}^{j} \) requires first estimating the output demand system using plant-product output and prices. When the resulting price elasticity matrix proves invertible, researchers derive \( MC_{it}^{j} \) under explicit market conduct assumptions, such as static Bertrand-Nash competition.

In his empirical application, \citet{orr2022within} employs a nested logit demand system\footnote{As shown in equation (20) of the original paper.}:  
\begin{equation}
    rs_{it}^{j} - rs_{t}^{0} = (1-\sigma) rs_{it}^{j|g(j)} - \alpha p_{it}^{j} + \eta_{it}^{j}, \label{eq:nested_logit}
\end{equation}
where \( rs_{it}^{j} = \ln{\frac{R_{it}^{j}}{I_{t}^{h(j)}}} \), with \( R_{it}^{j} \) denoting revenue from product \( j \) of plant \( i \) at time \( t \) and \( I_{t}^{h(j)} \) capturing total revenue of the 3-digit ASICC code \( h(j) \). The term \( rs_{it}^{j|g(j)} = \ln{\frac{R_{it}^{j}}{\Lambda_{t}^{g(j)}}} \), where \( \Lambda_{t}^{g(j)} \) denotes total revenue of the 5-digit ASICC code \( g(j) \). The variable \( p_{it}^{j} \) captures the logged output price for product \( j \), while \( \eta_{it}^{j} \) reflects product appeal. The price elasticity matrix proves invertible if and only if \( \alpha > 0 \) and \( 0 < \sigma < 1 \).

\section{Computational Reproducibility}\label{s:comp}

Using \href{https://www.journals.uchicago.edu/doi/suppl/10.1086/720465}{the replication package} from \citet{orr2022within}\footnote{
Our replication process identifies an issue in the provided package, specifically in \texttt{asicc\_code\_cleaning.do}, where line 130 excludes plants in industries 74 and 75 due to missing three-digit codes in the MOSPI ASICC09 file. We revise the do file to address this discrepancy and successfully reproduce all tables and figures from \citet{orr2022within}. We thank Scott Orr for clarifying the ASICC09 file structure and providing a modified version compatible with the original code.
}, we reproduce all main results.\footnote{Results of this reproduction exercise appear in Online Appendix, Section A.} The package provides data cleaning procedures while excluding raw datasets and analytical files. We obtain raw data for Indian manufacturing plants in the machinery sector (2000--2020) from the Ministry of Statistics and Programme Implementation and focus on the 2000--2007 sample, following the original study's methodology.

Our reproduction exercise successfully replicates the demand estimates from \citet{orr2022within} while revealing minor discrepancies in production function coefficients, detailed in Table \ref{tab:table_reproduction}. GMM estimation generates output elasticities of 0.617 for labor, 0.239 for capital, and 0.223 for materials, contrasting with the original preferred estimates of 0.626, 0.236, and 0.217 from column (3).\footnote{These deviations likely originate from computational differences in STATA's Mata matrix inversion functions across software versions and operating systems, which influence marginal cost calculations and subsequent input allocation decisions.} These minor variations preserve the main findings both qualitatively and quantitatively.

\begin{table}[ht]
   \centering
   \caption{Computational Reproduction of Columns 2, 3, and 4 in Table 4 of \cite{orr2022within}: Cobb-Douglas Production Function Estimates} \label{tab:table_reproduction}
      \resizebox{1\textwidth}{!}{ 
   \begin{threeparttable}
         \footnotesize 
      \input{Tables/table2_short}
        \begin{tablenotes}
        \footnotesize
        \item{Notes: Columns 1, 2, and 3 present computational reproductions (Rep.) of Columns 2, 3, and 4, respectively, from Table 4 of \cite{orr2022within} (Original Study). Calculations employ the 2000–2007 Indian ASI dataset provided by the Ministry of Statistics and Programme Implementation (MOSPI). Each observation represents a plant-product combination at the 5-digit ASICC variety level. The sample restricts analysis to producers within the machinery, equipment, and parts industry. The dependent variable captures the log quantity of product $j$ produced by plant $i$. Parentheses report plant-level block bootstrapped standard errors with 1,000 replications.
}
        \end{tablenotes}  
   \end{threeparttable}       
   }
\end{table}

\section{Robustness of Threshold for Instrument Construction}\label{s:IV}
\cite{orr2022within} excludes input codes utilized by single-product producers when machinery, equipment, and parts producers account for more than 30\% of revenue derived from purchases of these input codes. This exclusion criterion preserves instrument validity. As \cite{orr2022within} argues, ``input price variation should be driven by demand and supply shocks in other industries that are orthogonal to machinery demand shocks or general changes in machinery quality by product code. This requires that changes in average input prices not be driven by machinery demand.''

\begin{figure}[h]
    \centering
    \caption{Sensitivity of Demand Estimates Across Instrument Construction Thresholds}
    \label{fig:main_IV_thres}
    \begin{subfigure}[t]{0.495\textwidth}
        \centering
        \includegraphics[width=\linewidth]{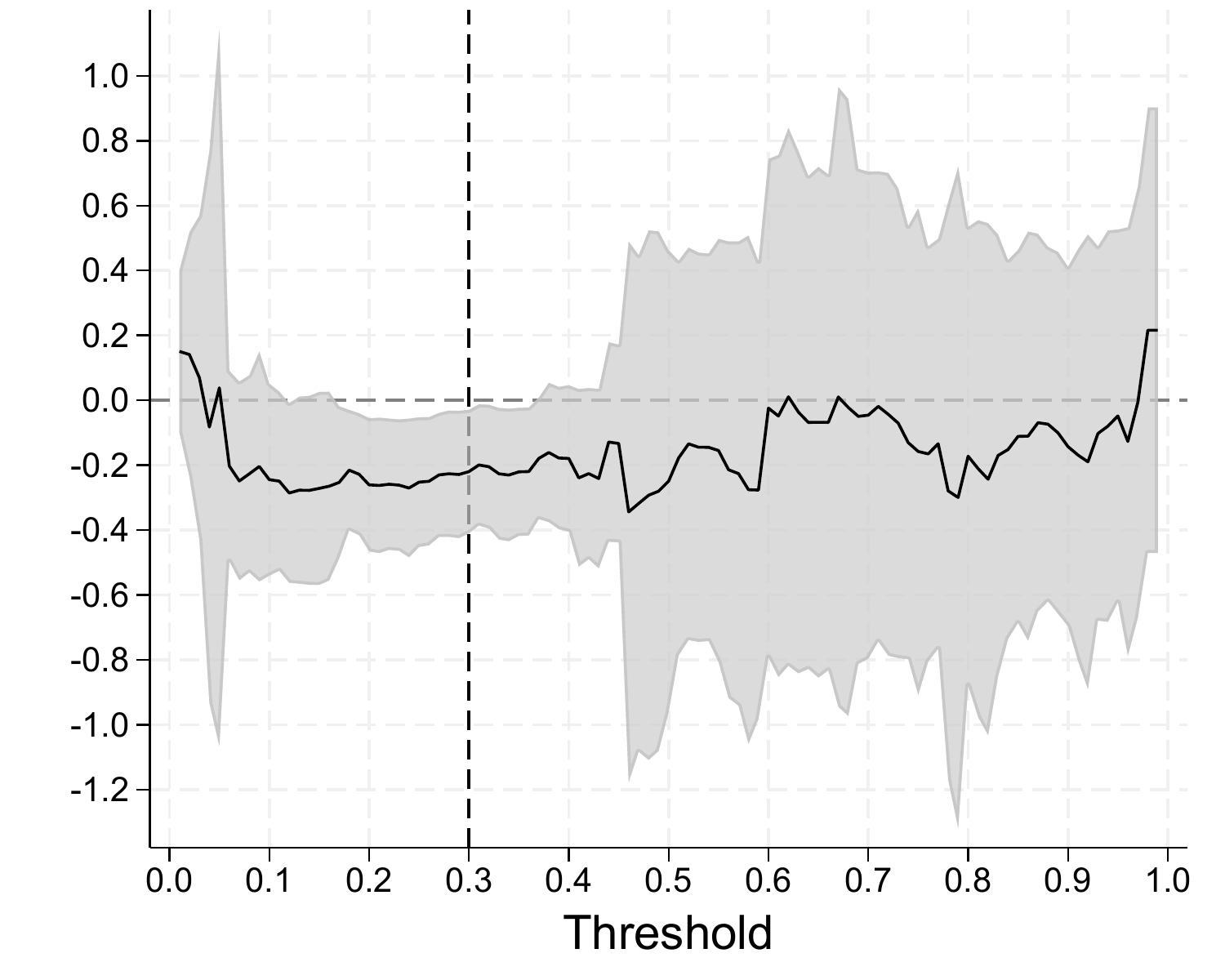}
        \caption{$p^j_{it}$ IV Coefficient}
        \label{fig:p_IV}
    \end{subfigure}
    \hfill
    \begin{subfigure}[t]{0.495\textwidth}
        \centering
        \includegraphics[width=\linewidth]{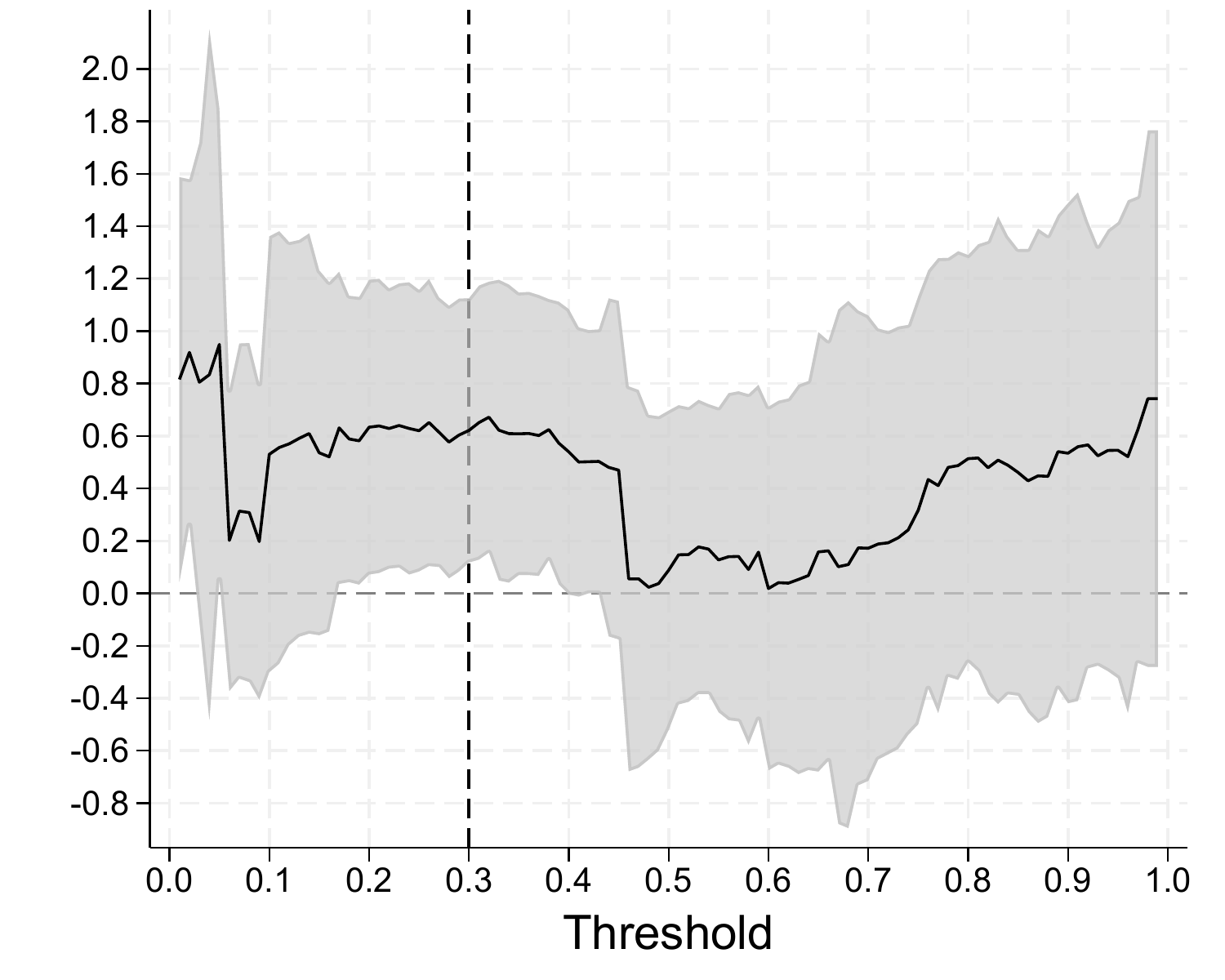}
        \caption{$rs^{j|g(j)}_{it}$ IV Coefficient}
        \label{fig:rs_IV}
    \end{subfigure}

    \vspace{0.5cm}  

    \begin{subfigure}[t]{0.495\textwidth}
        \centering
        \includegraphics[width=\linewidth]{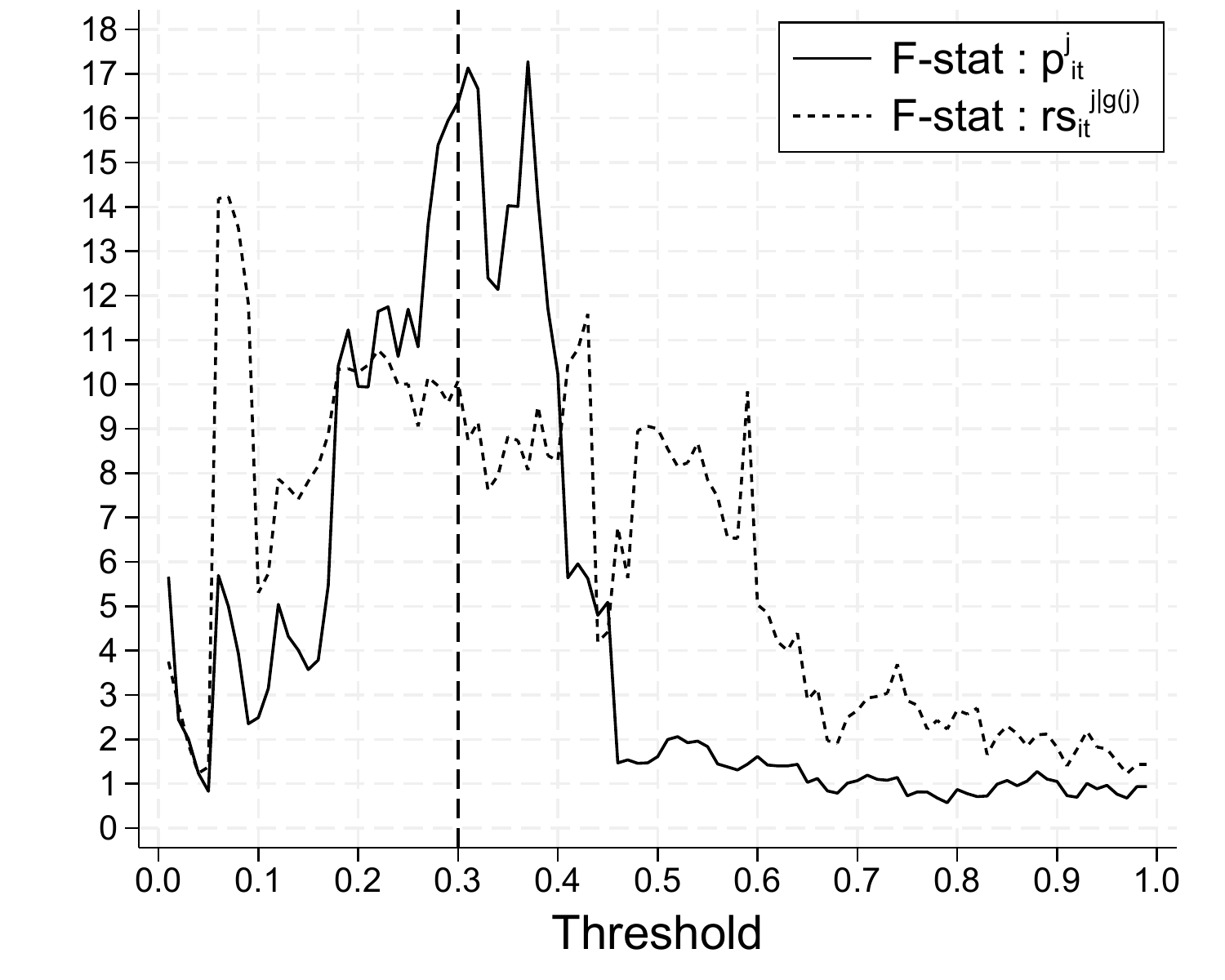}
        \caption{Sanderson-Windmeijer F-statistics}
        \label{fig:F_IV}
    \end{subfigure}
    \hfill
    \begin{subfigure}[t]{0.495\textwidth}
        \centering
        \includegraphics[width=\linewidth]{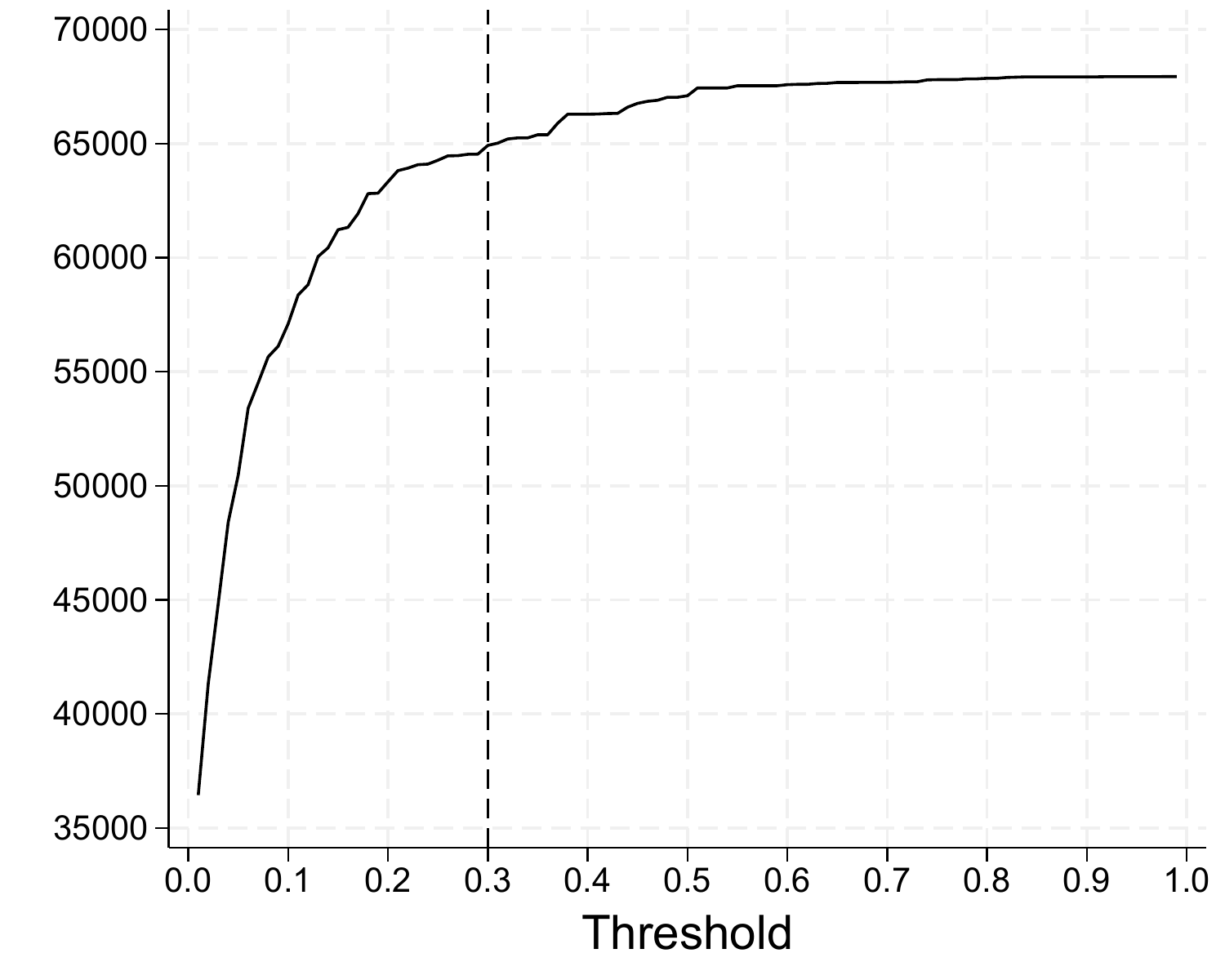}
        \caption{Observations}
        \label{fig:IV_obs}
    \end{subfigure}
        \begin{tablenotes}
        \scriptsize 
        \item{\item Notes: 
       This figure displays changes in IV estimates reported in Table 3 of \cite{orr2022within} as the instrument construction threshold varies in 0.01 increments. The y-axis presents distinct measures across panels: Panel (a) shows the estimated IV coefficient for $p^j_{it}$, Panel (b) displays the estimated IV coefficient for $rs^{j|g(j)}_{it}$, Panel (c) reports Sanderson-Windmeijer first-stage F-statistics for $p^j_{it}$ and $rs^{j|g(j)}_{it}$, and Panel (d) indicates the number of observations employed. Panels (a) and (b) include 90\% confidence intervals constructed using robust standard errors clustered by plant and product. A vertical dashed line marks the threshold employed in \cite{orr2022within} (0.3). The calculations utilize the 2000--2007 Indian Annual Survey of Industries (ASI) dataset from the Ministry of Statistics and Programme Implementation (MOSPI). 
        }
        \end{tablenotes}  
\end{figure}

The IV construction threshold substantially affects demand estimates. We replicate the analysis across thresholds ranging from 0 to 100\% in 1\% increments and present results in Figure \ref{fig:main_IV_thres}. First, the threshold directly determines sample size: lower thresholds reduce observations while higher thresholds retain more data but increase endogeneity risk. Panel \ref{fig:IV_obs} displays this variation in sample size.

Second, Panels \ref{fig:p_IV} and \ref{fig:rs_IV} display the price and revenue share parameter estimates. Excluding boundary cases near 0\% and 100\%, the price coefficient consistently yields a downward-sloping demand curve (ranging from -0.34 to -0.01), while the revenue share coefficient lies within the theoretical unit interval (0.02 to 0.95). Statistical significance varies markedly across thresholds, achieving 90\% confidence only within the 20\%--40\% range.

Finally, Panel \ref{fig:F_IV} documents instrument strength using Sanderson-Windmeijer F-statistics. Price instruments attain optimal performance within the 20\%-40\% threshold range, yielding F-statistics between 10 and 17. Outside this interval, F-statistics deteriorate to 1-6. Revenue share instruments demonstrate higher stability, sustaining F-statistics of 8-11 throughout the 20\%-40\% threshold range. Online Appendix B, Figure A3 presents additional first-stage coefficient estimates.

\begin{figure}[h]
    \centering
    \caption{Sensitivity of Efficiency Gains and Product Dropping Probabilities Across Instrument Construction Thresholds}
    \label{fig:main_RES_thres}
    \begin{subfigure}[t]{0.495\textwidth}
        \centering
        \includegraphics[width=\linewidth]{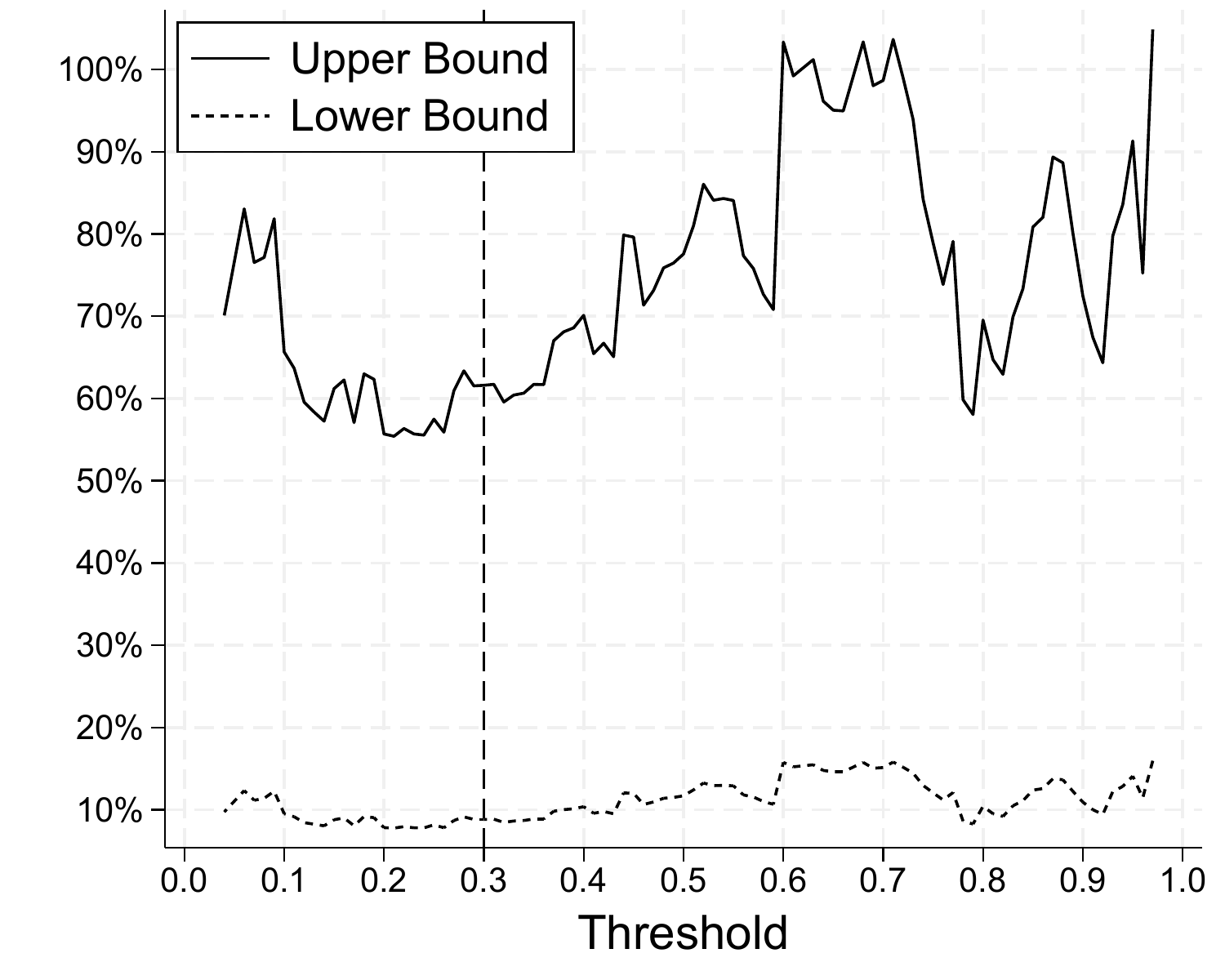}
        \caption{Plant-level Efficiency Growth}
        \label{fig:Eff}
    \end{subfigure}
    \hfill
    \begin{subfigure}[t]{0.495\textwidth}
        \centering
        \includegraphics[width=\linewidth]{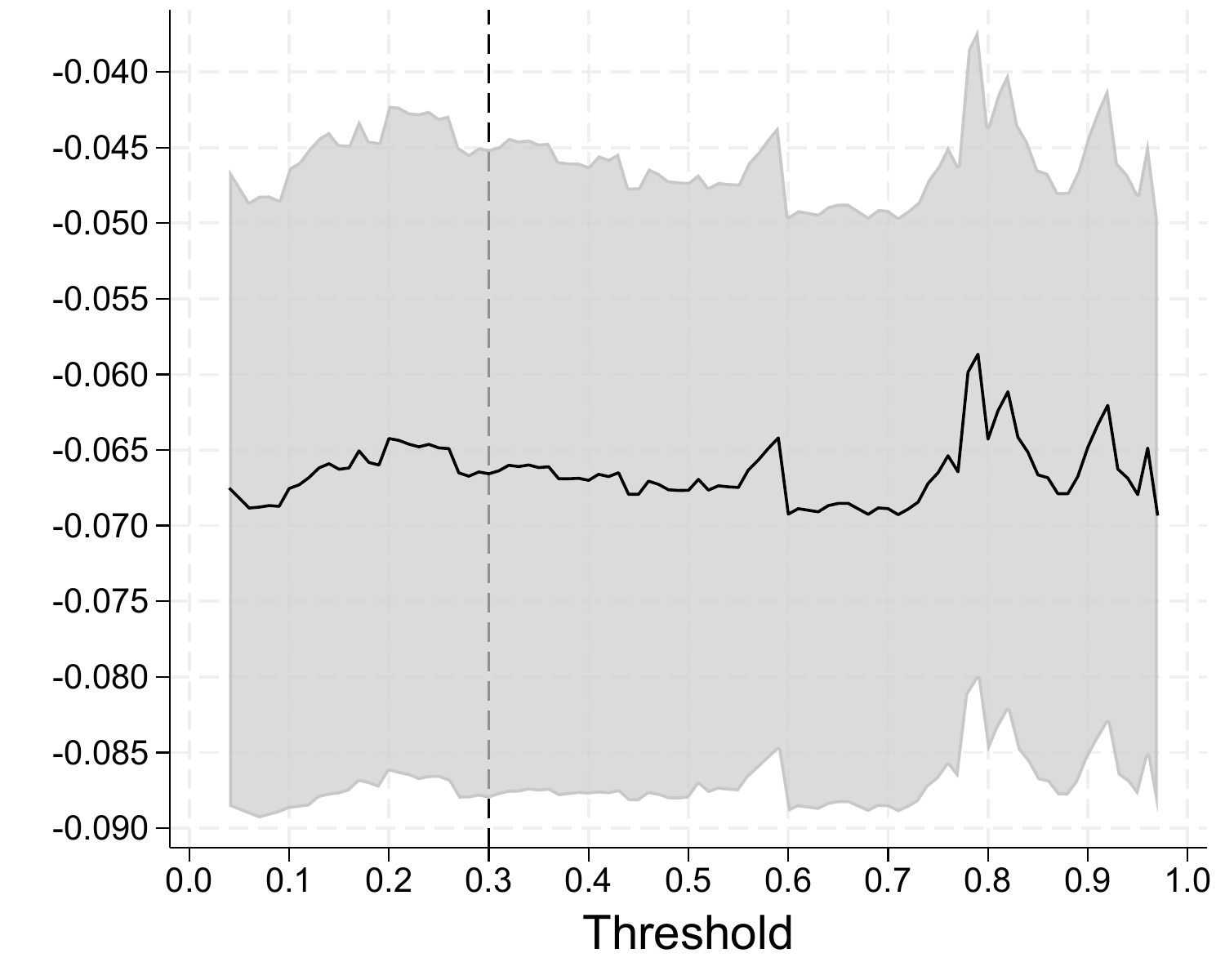}
        \caption{Product-Dropping Probit Estimates}
        \label{fig:Probit}
    \end{subfigure}
        \begin{tablenotes}
        \scriptsize 
        \item{\item Notes: 
       This figure presents robustness checks for the efficiency gains and product dropping probability estimates from Figure 2 and Table 9, column (1) of \cite{orr2022within}, examining how these results vary as the instrument construction threshold changes in 0.01 increments. Panel (a) displays the upper and lower bounds of plant-level efficiency growth following marginal variety removal for plants producing 2--10 products, estimated using the nonparametric specification. Panel (b) presents the marginal effect from a probit regression that predicts the probability of product discontinuation in the subsequent period as a function of the product's revenue efficiency, with marginal effects calculated at the sample means of all covariates. Panel (b) includes 90\% confidence intervals constructed using cluster-robust standard errors for the average marginal effect. The vertical dashed line indicates the threshold value of 0.3 employed in the original analysis. All calculations draw from the 2000--2007 Indian Annual Survey of Industries (ASI) dataset compiled by the Ministry of Statistics and Programme Implementation (MOSPI). 
        }
        \end{tablenotes}  
\end{figure}

Given our estimation of downward-sloping demand curves across nearly the entire threshold interval, we examine result sensitivity regarding plant-level efficiency gains from marginal variety elimination and discontinuation probabilities following product-level productivity declines. Figure \ref{fig:main_RES_thres} demonstrates that these findings exhibit remarkable robustness across threshold values.

Under the nonparametric specification from \cite{orr2022within}, Panel \ref{fig:Eff} documents that plants discontinuing their worst-performing products realize plant-level revenue efficiency gains spanning 10--15\% at lower bounds and reaching 60--80\% at upper bounds, with peaks approaching 100\% for thresholds of 60--70\%. Panel \ref{fig:Probit} confirms that one standard deviation declines in product-level revenue-based total factor productivity increase discontinuation probabilities by 6--7 percentage points across all threshold values.\footnote{More detailed results are reported in Figures A1-A2 and Tables A3 in Online Appendix B.}

\section{Replication Using Different Sample Periods}\label{s:post}

\begin{figure}[h]
    \centering
    \caption{Sensitivity of Demand Estimates Across Instrument Construction Thresholds \\ Sample Period: 2010-2019}
    \label{fig:main_IV_thres_1019}
    \begin{subfigure}[t]{0.495\textwidth}
        \centering
        \includegraphics[width=\linewidth]{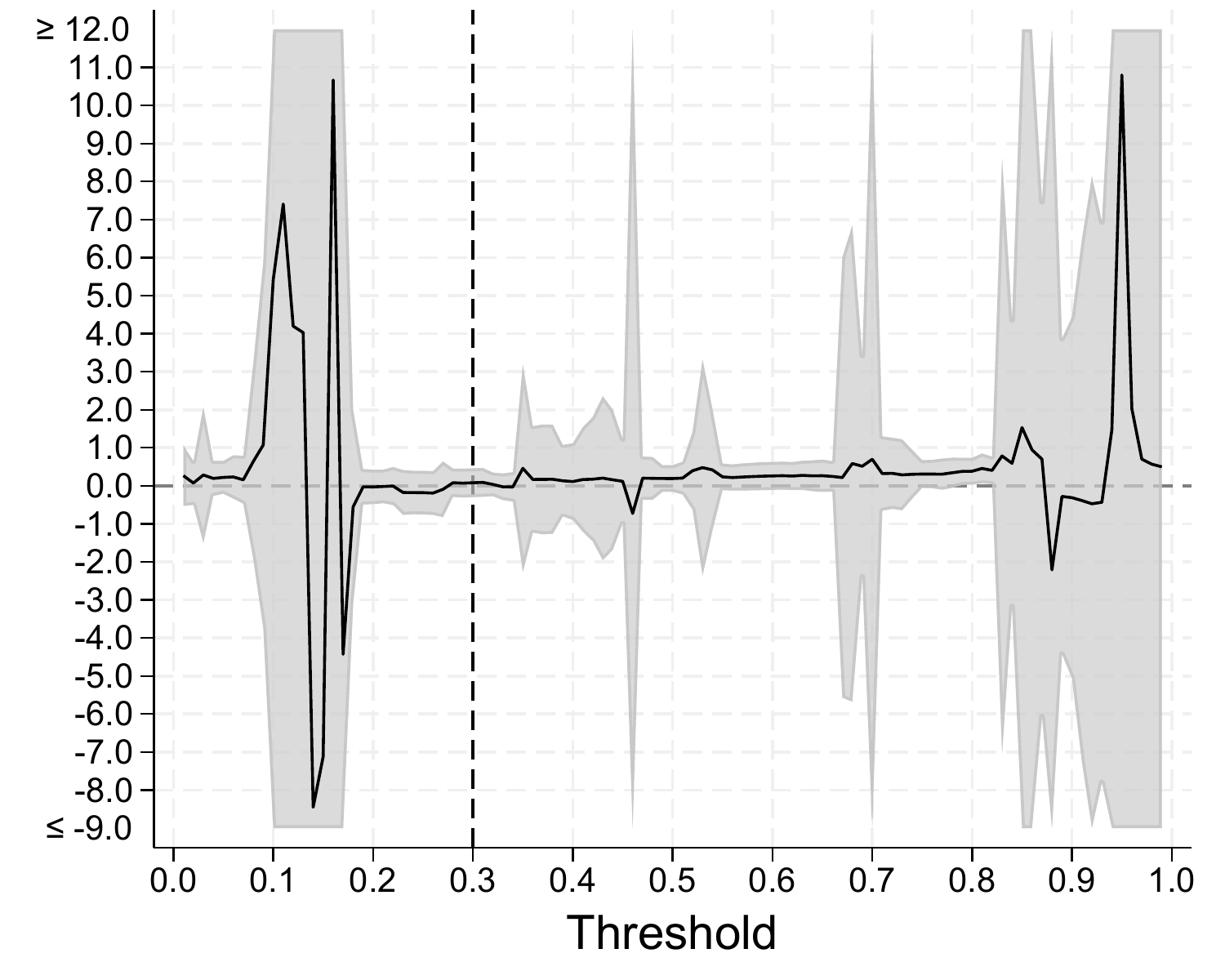}
        \caption{$p^j_{it}$ IV Coefficient}
        \label{fig:p_IV_1019}
    \end{subfigure}
    \hfill
    \begin{subfigure}[t]{0.495\textwidth}
        \centering
        \includegraphics[width=\linewidth]{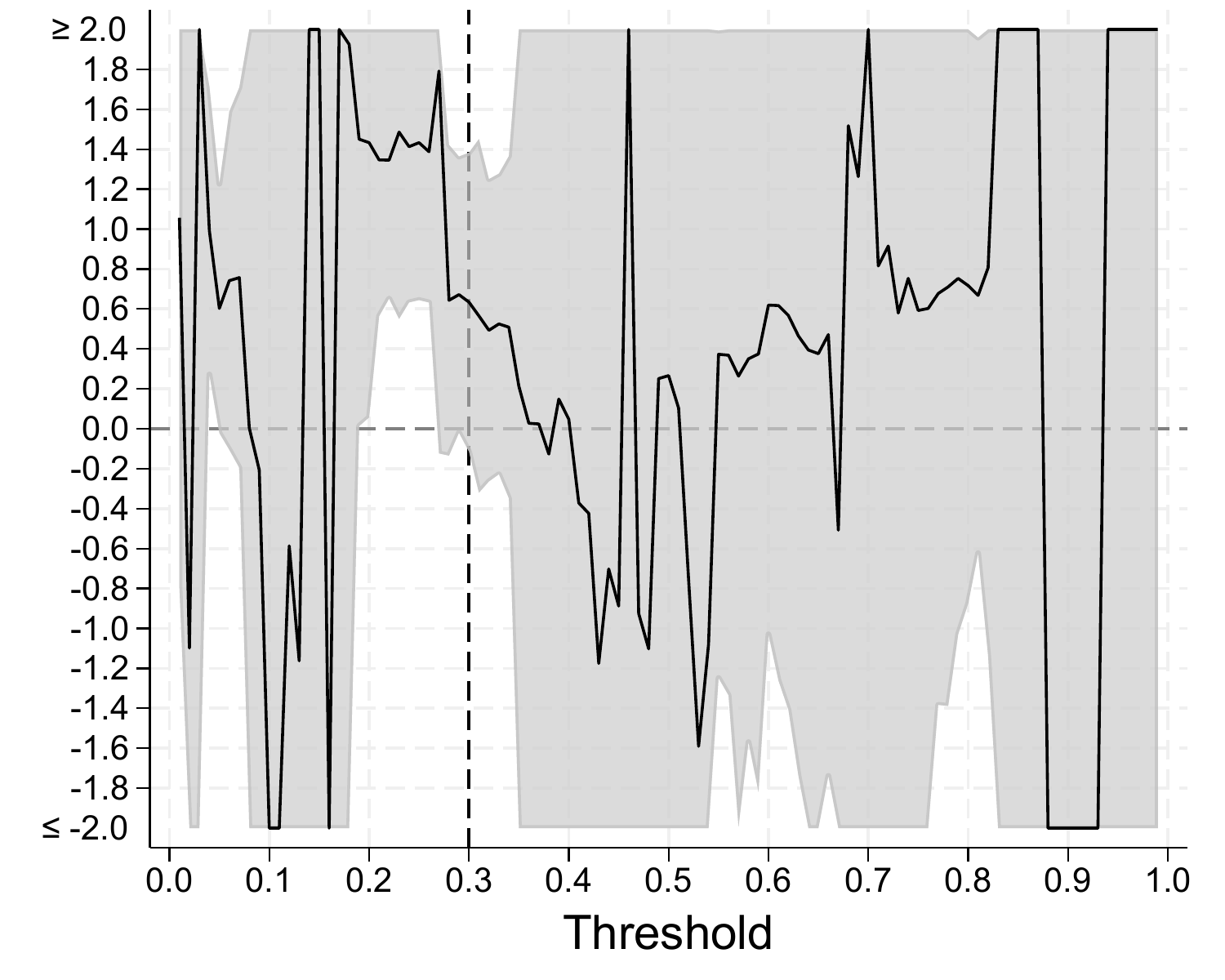}
        \caption{$rs^{j|g(j)}_{it}$ IV Coefficient}
        \label{fig:rs_IV_1019}
    \end{subfigure}

    \vspace{0.5cm}  

    \begin{subfigure}[t]{0.495\textwidth}
        \centering
        \includegraphics[width=\linewidth]{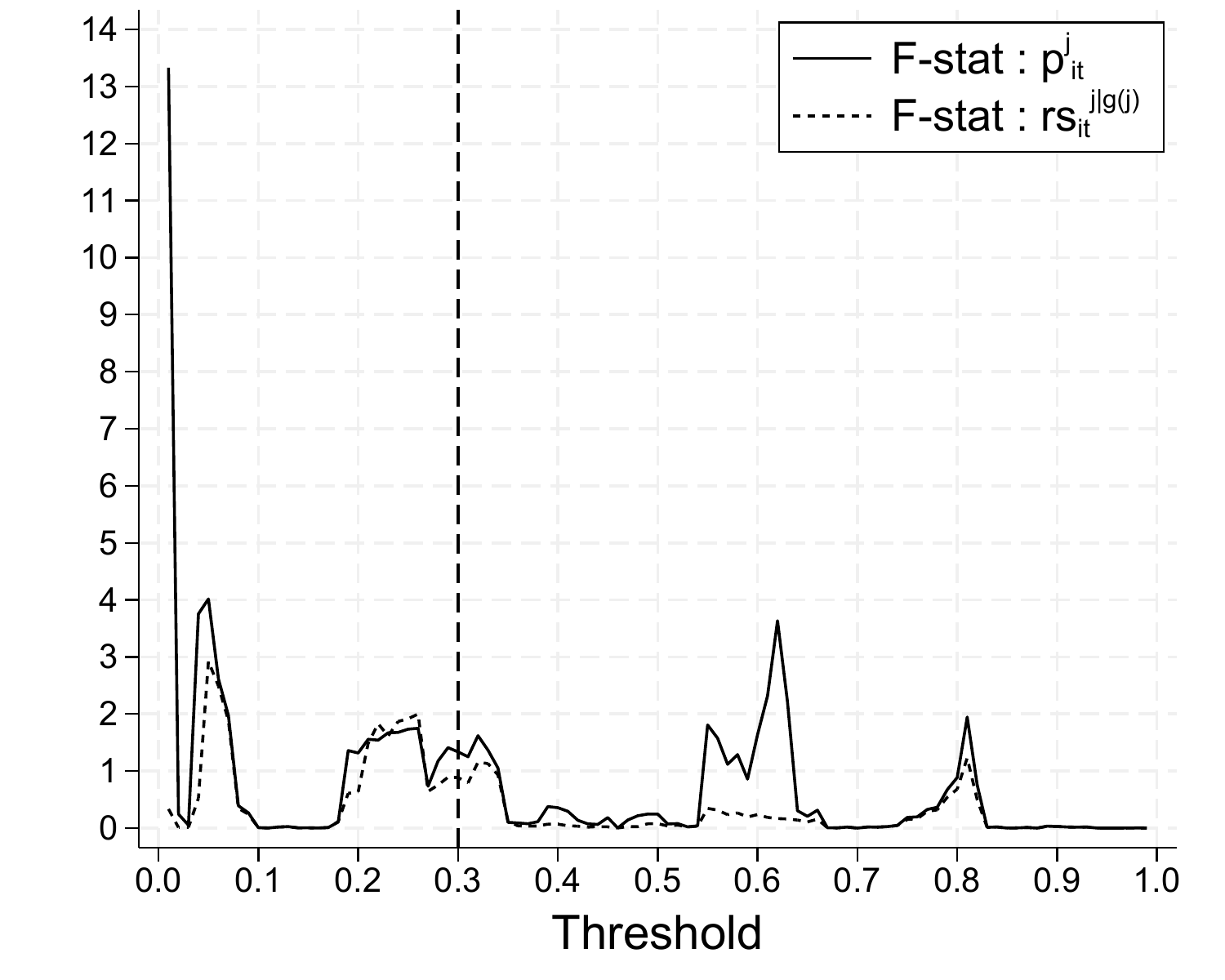}
        \caption{Sanderson-Windmeijer F-statistics}
        \label{fig:F_IV_1019}
    \end{subfigure}
    \hfill
    \begin{subfigure}[t]{0.495\textwidth}
        \centering
        \includegraphics[width=\linewidth]{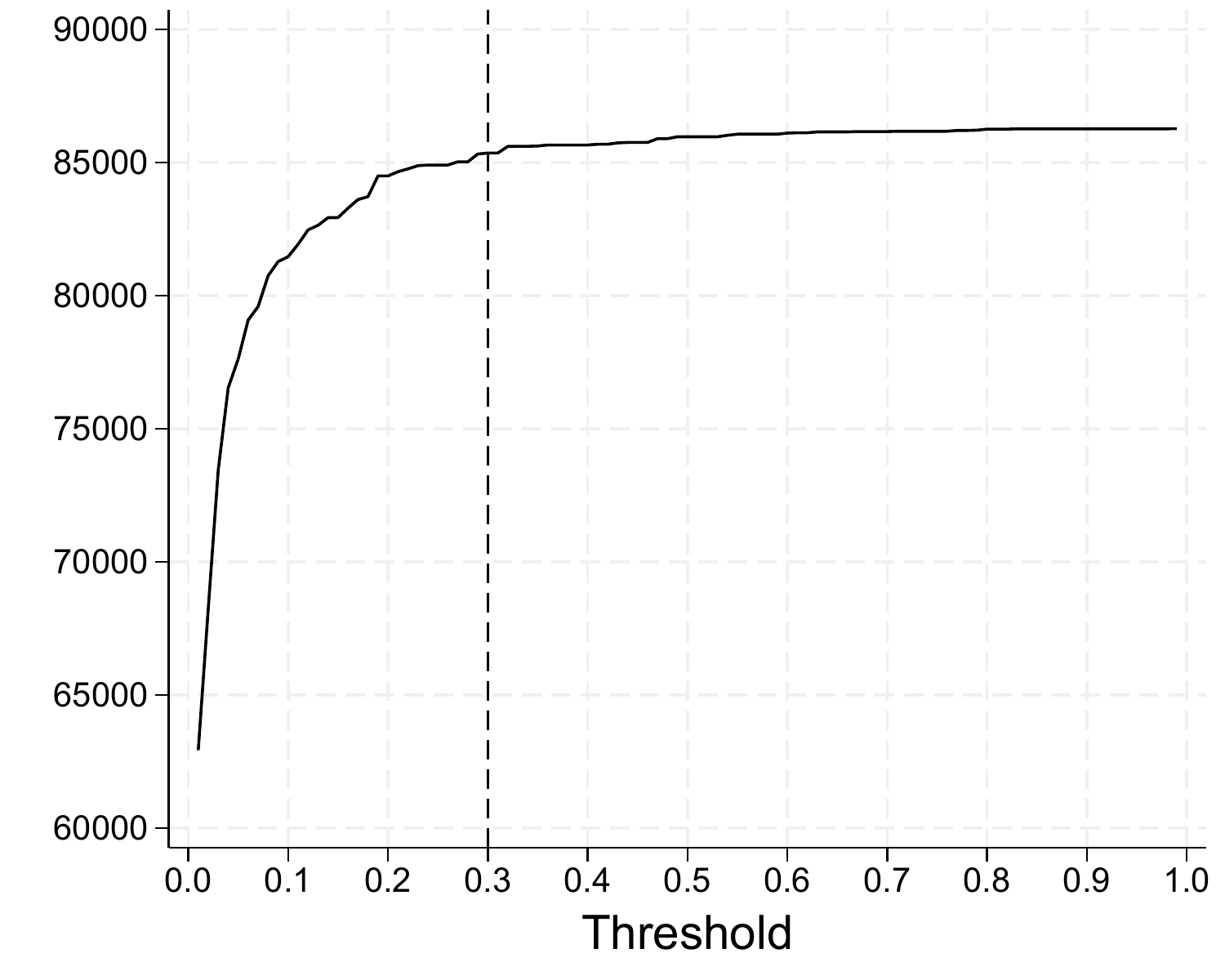}
        \caption{Observations}
        \label{fig:IV_obs_1019}
    \end{subfigure}
        \begin{tablenotes}
        \scriptsize 
        \item{\item Notes: 
       This figure displays changes in IV estimates reported in Table 3 of \cite{orr2022within} as the instrument construction threshold varies in 0.01 increments. The y-axis presents distinct measures across panels: Panel (a) shows the estimated IV coefficient for $p^j_{it}$, Panel (b) displays the estimated IV coefficient for $rs^{j|g(j)}_{it}$, Panel (c) reports Sanderson-Windmeijer first-stage F-statistics for $p^j_{it}$ and $rs^{j|g(j)}_{it}$, and Panel (d) indicates the number of observations employed. Panels (a) and (b) include 90\% confidence intervals constructed using robust standard errors clustered by plant and product. We bound the y-axis of panels (a) and (b) to [-9,12] and [-2,2], respectively, to enhance readability. A vertical dashed line marks the threshold employed in \cite{orr2022within} (0.3). The calculations utilize the 2010--2019 Indian Annual Survey of Industries (ASI) dataset from the Ministry of Statistics and Programme Implementation (MOSPI). 
        }
        \end{tablenotes}  
\end{figure}

\begin{figure}[h]
    \centering
    \caption{Sensitivity of Demand Estimates Across Instrument Construction Thresholds \\ Sample Period: 2000-2019}
    \label{fig:main_IV_thres_0019}
    \begin{subfigure}[t]{0.495\textwidth}
        \centering
        \includegraphics[width=\linewidth]{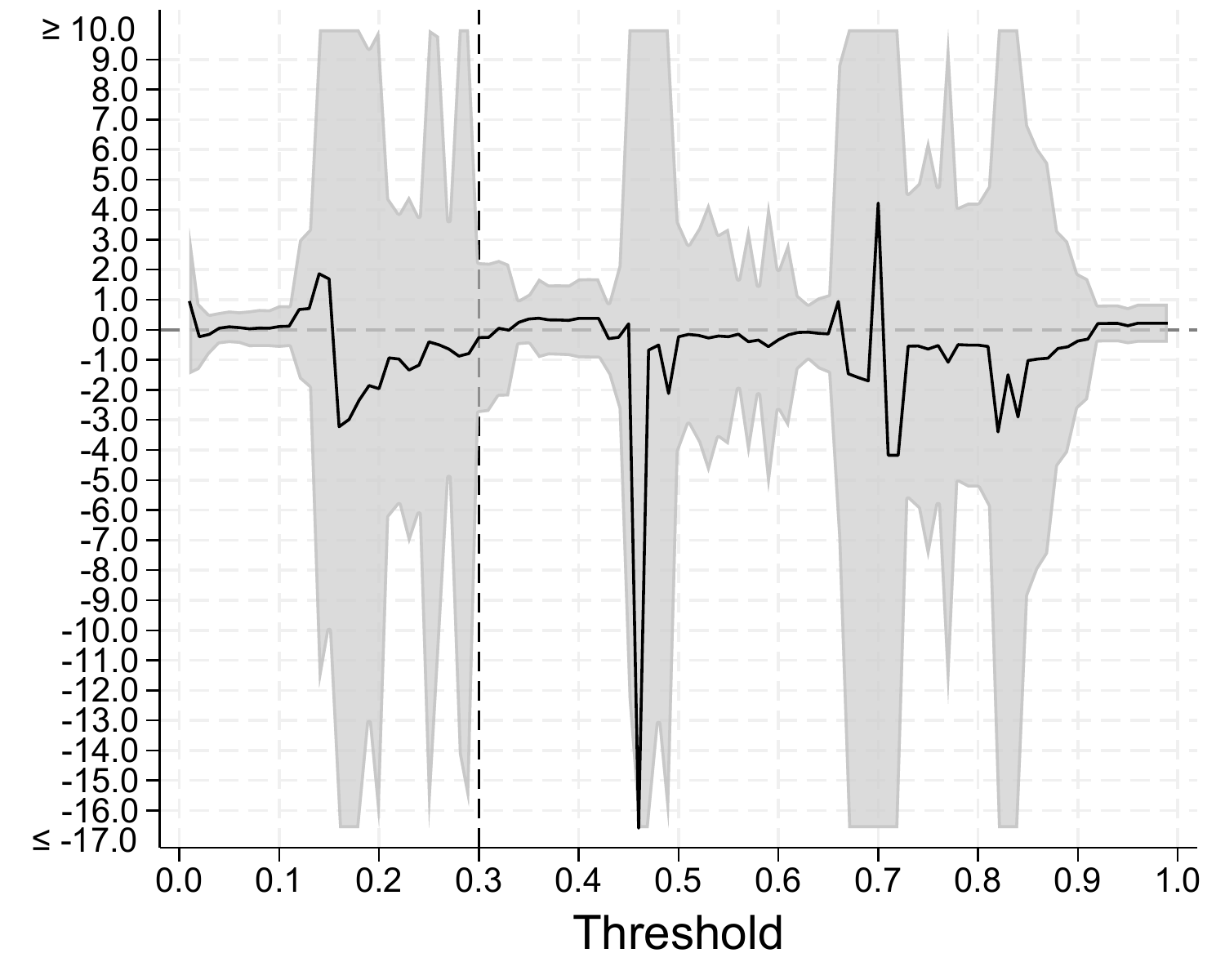}
        \caption{$p^j_{it}$ IV Coefficient}
        \label{fig:p_IV_0019}
    \end{subfigure}
    \hfill
    \begin{subfigure}[t]{0.495\textwidth}
        \centering
        \includegraphics[width=\linewidth]{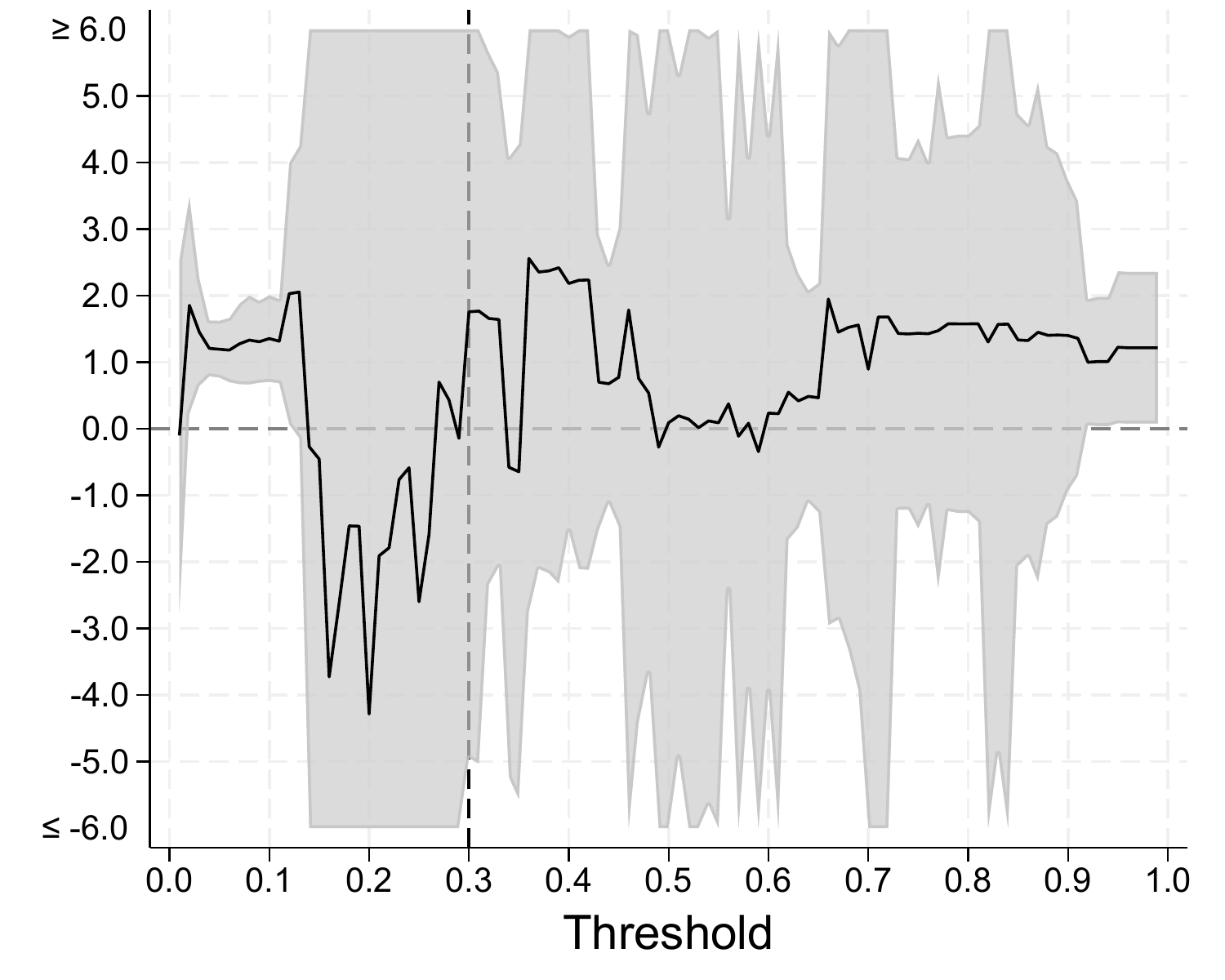}
        \caption{$rs^{j|g(j)}_{it}$ IV Coefficient}
        \label{fig:rs_IV_0019}
    \end{subfigure}

    \vspace{0.5cm}  

    \begin{subfigure}[t]{0.495\textwidth}
        \centering
        \includegraphics[width=\linewidth]{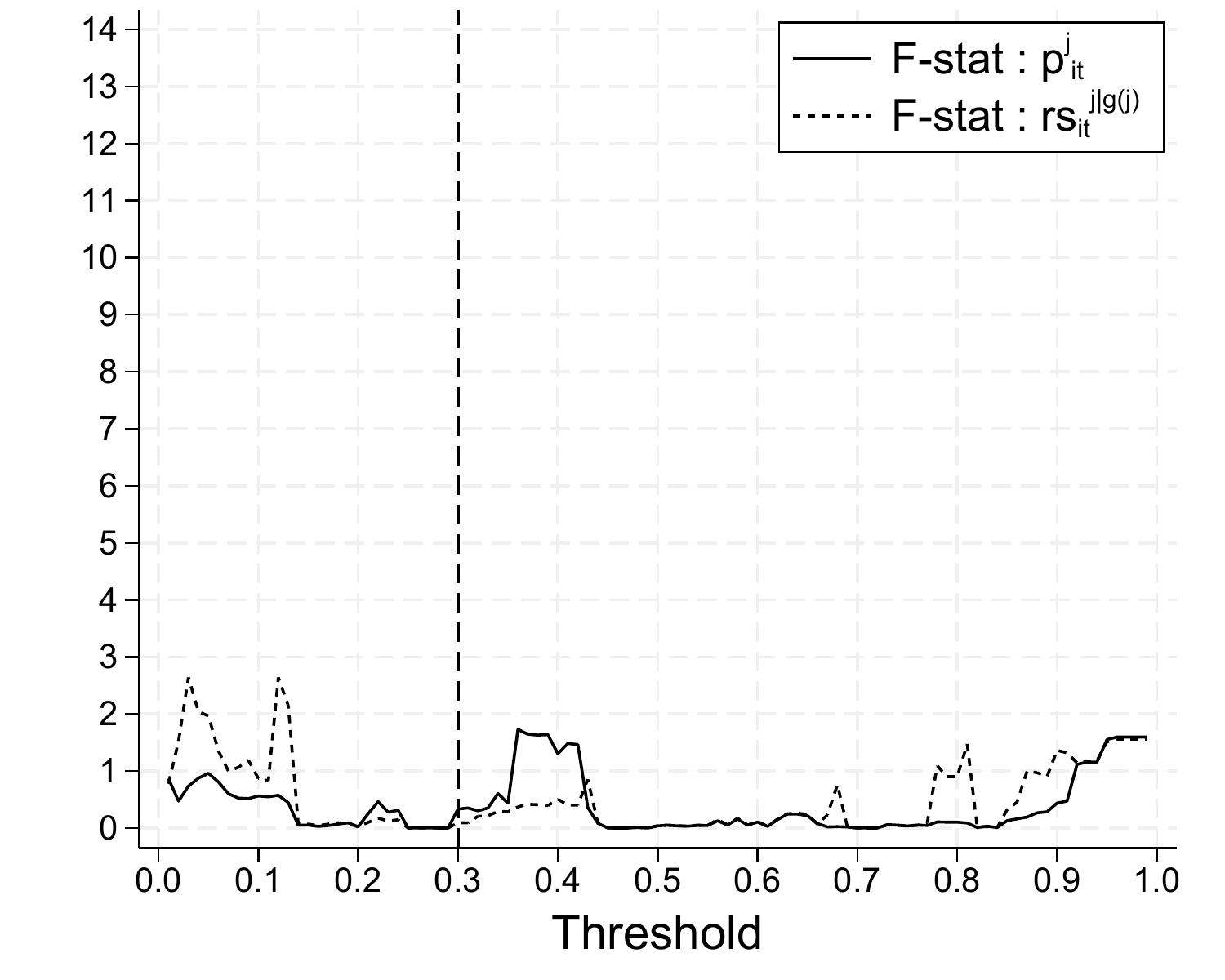}
        \caption{Sanderson-Windmeijer F-statistics}
        \label{fig:F_IV_0019}
    \end{subfigure}
    \hfill
    \begin{subfigure}[t]{0.495\textwidth}
        \centering
        \includegraphics[width=\linewidth]{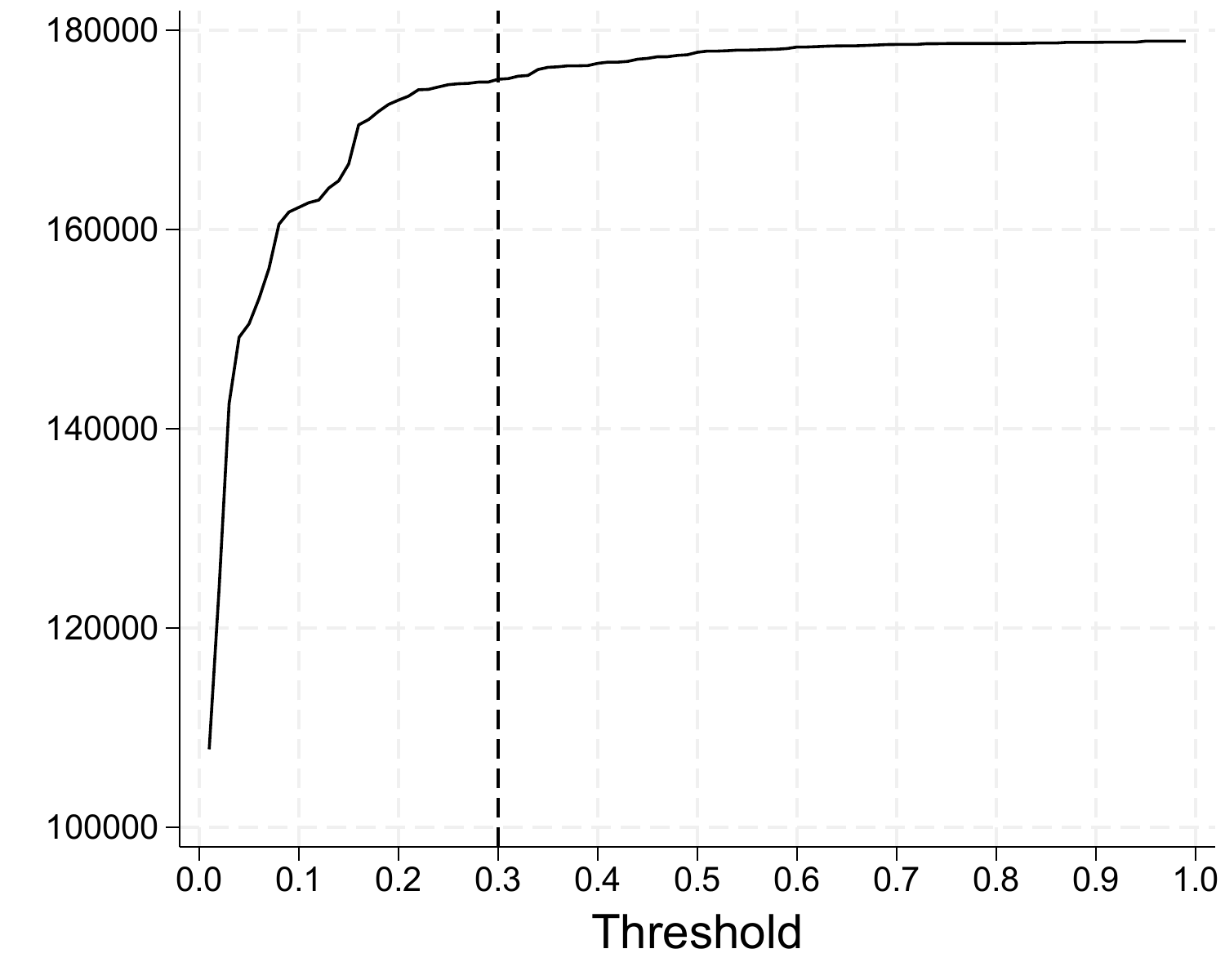}
        \caption{Observations}
        \label{fig:IV_obs_0019}
    \end{subfigure}
        \begin{tablenotes}
        \scriptsize 
        \item{\item Notes: 
       This figure displays changes in IV estimates reported in Table 3 of \cite{orr2022within} as the instrument construction threshold varies in 0.01 increments. The y-axis presents distinct measures across panels: Panel (a) shows the estimated IV coefficient for $p^j_{it}$, Panel (b) displays the estimated IV coefficient for $rs^{j|g(j)}_{it}$, Panel (c) reports Sanderson-Windmeijer first-stage F-statistics for $p^j_{it}$ and $rs^{j|g(j)}_{it}$, and Panel (d) indicates the number of observations employed. Panels (a) and (b) include 90\% confidence intervals constructed using robust standard errors clustered by plant and product. We bound the y-axis of panels (a) and (b) to [-17,10] and [-6,6], respectively, to enhance readability. A vertical dashed line marks the threshold employed in \cite{orr2022within} (0.3). The calculations utilize the 2000--2019 Indian Annual Survey of Industries (ASI) dataset from the Ministry of Statistics and Programme Implementation (MOSPI). 
        }
        \end{tablenotes}  
\end{figure}

We apply the methodology of \citet{orr2022within} to (1) the ASI sample for 2010–2019 and (2) the full sample spanning 2000–2019.\footnote{For comparability, we convert post-2010 product codes (NPCMS) to ASICC09 using the official ASICC09–NPCMS concordance from MOSPI. We report summary statistics for the two samples in Online Appendix C.} We begin by evaluating the performance of the proposed IVs but consistently fail to obtain admissible estimates across the full range of threshold values. Figures \ref{fig:main_IV_thres_1019} and \ref{fig:main_IV_thres_0019} present the demand estimates for the 2010--2019 and 2000--2019 samples, respectively, along with first-stage F-statistics and the number of observations at each threshold. In both samples, estimated price and nest parameters lie outside the invertible region across virtually all thresholds. The price coefficient ranges from -8.45 to 10.80 (2010--2019) and -16.58 to 4.22 (2000--2019); the nest parameter from -16.10 to 147.83 and -4.28 to 2.56, respectively.\footnote{These extreme values are not visible in Panel \ref{fig:rs_IV_1019}, since the y-axis is bounded to [-2,2].} Moreover, our estimates yield no statistically significant results. Consistent with these findings, the instruments exhibit weak power throughout, with first-stage F-statistics never exceeding the conventional threshold of 10.

Although demand estimates from different periods violate the invertibility condition, we assess the robustness of \cite{orr2022within}'s main findings by recovering within-firm input allocation shares using the original demand estimates and estimating the production function.\footnote{Production function parameter estimates appear in Online Appendix Section C, Tables A5 and A6.}  We then quantify efficiency gains from eliminating the least productive products and evaluate the marginal effect of a one standard deviation decline in revenue-based productivity on product discontinuation probability.

Table \ref{tab:table_key} presents the main results from this robustness exercise. The patterns exhibit remarkable consistency across sample periods. Removing the lowest-performing products generates substantial increases in plant-level revenue-based productivity: 8.82\%–61.60\% for the original sample, 1.13\%–67.13\% for 2011–2020, and 9.94\%–65.59\% for 2001–2020. The marginal effects of a one standard deviation decline in revenue-based productivity on the probability of product discontinuation display similar stability across periods, with magnitudes of 6.65, 9.22, and 7.96 percentage points, respectively.\footnote{For more detailed results, please refer to Figures A7-A10 and Tables A7 and A8 in Online Appendix C.}

\begin{table}[H]
   \centering
   \caption{Baseline Within-Firm Productivity Dispersion Implications} \label{tab:table_key}
   \begin{threeparttable}
         \footnotesize 
      \input{Tables/table_keynumbers} 
        \begin{tablenotes}
        \scriptsize 
        \item{Notes: This table documents robustness checks on the baseline within-firm productivity dispersion findings from \citet{orr2022within}. The first column presents lower and upper bounds of plant-level efficiency gains from eliminating marginal varieties of lowest-performing products, alongside marginal effects of one standard deviation declines in revenue-based productivity on product exit probabilities, employing 2000–2007 ASI data from India's Ministry of Statistics and Programme Implementation (MOSPI). Columns two and three replicate this analysis using 2010–2019 and 2000–2019 ASI samples, respectively. Parentheses report plant-level semi-parametric block bootstrap standard errors based on the parametric distribution of estimated demand parameters from the original sample (1,000 replications).
}
        \end{tablenotes}  
   \end{threeparttable}   
\end{table}

\section{Conclusion}\label{s:conc}

We evaluate the replicability of the methodology proposed by \cite{orr2022within} for estimating within-plant productivity across product lines. We reproduce the original results and document that demand estimates are sensitive to the thresholds used in constructing instrumental variables. When we extend the analysis to alternative sample periods (2010--2019 and 2000--2019), the demand estimation becomes unstable, requiring calibration of key parameters from the original sample. Nonetheless, the study's main findings prove robust: the estimated efficiency gains from discontinuing underperforming products, as well as the relationship between productivity declines and product exit probabilities, remain consistent across sample periods.

\clearpage

\thispagestyle{empty}

\bibliographystyle{kluwer}
\bibliography{biblio}

\clearpage

\section{Figures}

\setcounter{page}{9}


\clearpage

\end{spacing}

\end{document}

%% file: Tables/table2_short.tex
\begin{tabular}{lcccccc} 
\hline\hline    
  &      \multicolumn{6}{c}{GMM} \\
  &      \multicolumn{2}{c}{(1)} & \multicolumn{2}{c}{(2)} & \multicolumn{2}{c}{(3)} \\
  &      Original &  Rep.  &  Original& Rep.   &   Original   & Rep.   \\
  &      Study   &       &  Study&      &   Study   &    \\
\hline
$\beta_{L}$ &     0.331      &  0.325     &    0.321  &   0.315    &  0.626         &     0.617 \\
            &    (0.192)   &  (0.192)    &   (0.191)   & (0.191)    & (0.261)       &   (0.261)   
\\
$\beta_{K}$ &     0.101      &   0.106    &    0.097        &    0.102   &      0.236    &   0.239   \\
            &    (0.082)     &  (0.082)     &   (0.081)    &   (0.081)    &   (0.099)     &  (0.099) 
 \\           
$\beta_{M}$ &      0.790    &  0.789     &   0.806    &    0.806   &  0.217        &    0.223  \\
            &     (0.191)    &   (0.191)    &   (0.186)  &  (0.186)     &  (0.352)     &  (0.352)   
\\                  
\hline
\multicolumn{6}{l}{\textbf{Ins.}}  \\                  
$(Z_{t}^{g(j)}, Z_{t-1}^{g(j)})$             &  $\checkmark$     &   $\checkmark$   &       &       &     $\checkmark$      &     $\checkmark$  \\
$m_{it-1}$  &   $\checkmark$      &    $\checkmark$    &    $\checkmark$    &     $\checkmark$   &          &      \\
\hline  
Observations    &   3,620  &  3,620  &  3,620  &  3,620  &  3,620  &  3,620 \\
\hline \hline
\end{tabular}

%% file: Tables/table_keynumbers.tex
\begin{tabular}{lccc}
\hline\hline    
     Period                            &      2000-2007  &  2010-2019 & 2000-2019 \\
\hline
Plant-level Efficiency Growths (\%) &     [8.82, 61.60]      &  [1.13, 67.13]     &   [9.94, 65.59]  \\
\\
Marginal Effect of 1 SD TFPR Decline   &     6.65      &  9.22     &    7.96  \\
 on Product-Dropping (pp)              &    (2.20)   &  (1.53)    &   (2.08)    \\
\hline
\end{tabular}